\documentstyle[rmaaconf]{article}

\def\et{et\thinspace al.\ }                                 %et al.%
\def\gapprox{_>\atop{^\sim}}       %Greater than over approximately%
\def\percc{cm$^{-3}$}                         %per cubic centimeter%
\def\Ha{$H\alpha$}                                         %H alpha%
\def\Hb{$H\beta$}                                           %H beta%

\begin{document}

\title{The Starburst Model for AGN: Past, Present \& Future}

\author{Roberto Cid Fernandes\altaffilmark{1}}

\altaffiltext{1}{Depto.\ de F\'{\i}sica, CFM - UFSC, Campus
Universit\'ario - Trindade, 88040-900 ~~Florian\'opolis, SC, Brazil.\\
cid@fsc.ufsc.br, http://www.if.ufrgs.br/$\sim$cid}

    \begin{resumen} 
    Han pasado once a\~nos desde que Terlevich y Melnick propusieron
por primera vez su modelo de AGN sin agujeros negros, una idea que
desde entonces ha evolucionado a lo que hoy conocemos como modelo de
starburst para gal\'axias ativas. Este modelo ha sido tema de
discusi\'on y pol\'emica en la \'ultima decada, con evidencias
observacionales tanto a favor como en contra. Despues de todos estos
a\~nos, ?`podemos llegar a un veredicto sobre si los starbursts pueden
ser responsables de la actividad en AGN? En esa contribuci\'on
intentamos responder a esa questi\'on, revisando las principales
virtudes y problemas del modelo.
    \end{resumen}

    \begin{abstract}
    It is now eleven years since Terlevich \& Melnick first proposed an
`AGN without black-holes' model, an idea which since then evolved into
what is now called the starburst model for AGN. This model has been the
subject of much debate in the last decade, with observational evidence
both for and against it further fuelling the controversy. Can we after
all these years reach a veredictum on whether starbursts can power AGN?
This contribution tries to answer this question reviewing the main
achievements of the starburst model, its current status and future
prospects.
    \end{abstract}

\section{Introduction}

Probably the most extreme example of `starburst activity in galaxies'
occurs when a nuclear starburst behaves like an AGN. But is that
possible? Can the basic properties of Active Galactic Nuclei be
understood in terms of physical processes associated with a starburst?

By the mid 80's the answer to this question was `no', as there seemed
to be no reasonable way to explain basic AGN properties such as
variability, emission line ratios and widths in terms of stellar
processes. The consensus then emerged that accretion onto a
super-massive black hole is the ultimate source of energy in active
galaxies, though it is fair to say that this consensus was at least
partially established due to the lack of plausible alternative
theories. Since then Terlevich and colaborators have given new breath
to starburst-based models for AGN exploring new possibilities such as
warmers and compact Supernova Remnants (cSNRs).

There have been several reviews of the starburst model for AGN, such
as those by Terlevich (1989, 1992, 1994), Filippenko (1992)
and Heckman (1991). This review differs from the previous ones
because it comes in a time when, curiously, though there have never
been as much evidence for a connection between star-formation and
activity, there are also new strong evidence for the existence of
accretion disks and black holes in AGN, as the reader will realize
browsing through this very volume. This may well be indicating that
the answer to the AGN phenomenum lies with neither  starbursts nor
black-holes, but {\it both}, an idea which gathered some strength
during this conference. It is however important to establish the
virtues and limitations of both theories before reaching such a
strong conclusion. In this paper I review the essential aspects of
the starburst model for AGN, discussing which aspects of the
activity phenomenum can be adequately understood in terms of the
physics of starbursts and cSNRs, as well as properties which do not
seem to fit in this framework.

\section{Overview of the model}

The starburst model started with the idea that hot stars could power
the ionization in narrow lined AGN, and gradually switched its focus
to the phenomena associated with cSNRs in a nuclear starburst. In
this section I briefly review the history model, indicating strong
and weak points and updating it where necessary.

\subsection{Warmers}

\label{sec:warmers}

The official kick-off of the starburst model was given in 1985 by
Terlevich \& Melnick. They proposed a scenario in which the central
ionizing source consists of a young, massive, metal-rich cluster
containing `warmers', extreme WR stars reaching temperatures of $\sim
150000$\ K. The existence of such stars was supported by stellar
evolution calculations (e.g., Maeder \& Meynet 1988) which showed that
strong mass-loss could strip-off the outer layers of massive stars,
exposing their hot interior. Terlevich \& Melnick's calculations (later
confirmed by Cid Fernandes \et 1992 and Garcia-Vargas \& D\'{\i}az
1992) showed that a cluster containing `warmers' would have an ionizing
spectrum hard enough to photoionize the surrounding gas to Narrow Line
Region like conditions.

Though a few warmers have been found (e.g., Dopita \et 1990), by now
it is clear that they are not as common as previously thought, most
likely because the mass lost during the evolution forms an opaque
atmosphere which effectively reduces the temperature of the star. In
fact, updated spectral synthesis calculations using newer
evolutionary tracks and appropriate stellar atmospheres for the WR
phases show that the ionizing spectrum is more typical of HII
regions than of AGN (e.g., Leitherer, Gruenwald \& Schmutz 1992). 

In summary, `warmers' did not live up to be the long sought `missing
link between starbursts and AGN'. This is not to say that WR-stars
have nothing to do with AGN. In fact, the existence of objects
showing both AGN and WR features points to some sort of connection
at least in some cases (see Conti 1993). It is nevertheless clear
that a sound starburst model for AGN could not be based on the
existence of warmers. Indeed, in the subsequent papers cSNRs
(\S\ref{sec:cSNRs}) have replaced warmers as the basic agent of
activity, to the point that warmers are {\it not} a necessary
ingreedient of the model anymore.

\subsection{Evolutionary Scheme}

\label{sec:Evol_Scheme}

Terlevich, Melnick \& Moles (1987) considered the evolution of a
metal-rich nuclear starburst, dividing it into four phases: (1) an
initial HII region phase lasting for $\sim 3$\ Myr, when all stars
are in the main-sequence, (2) a phase from 3 to 4 Myr when the first
warmers appear and the emission line spectrum changes to Seyfert 2
or LINER, (3) a phase from 4 to 8 Myr where warmers and OB stars
still dominate the ionization but the cluster also contains type Ib
SNe and red supergiants, and (4) a Seyfert 1 phase from 8 to 60 Myr,
when cSNRs produce variability, ionization and broad lines.

There are several reasons why this scheme should be revised, the
first of which is that, as discussed above, warmers can almost
certainly be ruled out as abundant sources of ionizing photons. 
Besides, we now know that, at least in some cases, the difference
between Seyfert 2s and Seyfert 1s is not an evolutionary one, but an
orientation effect. A revised evolutionary scheme would comprise
only two phases: an initial HII region (or LINER, depending on the
density and metallicity; Shields 1992, Filippenko \& Terlevich 1992)
phase lasting until the first cSNR appears at $\sim 8$~Myr and the
nucleus turns into a Seyfert 1. During both phases the action of
stellar winds and SNRs can drive a biconical outflow if the parent
molecular cloud (the torus?) is flattened. This mechanism would
naturally create the conditions postulated by the unified model.
Seyfert 2s would then be simply edge-on Seyfert 1s. Some Seyfert 2s
and LINERs could also be face-on low luminosity phase 2 systems,
where the low SN rate ensures quiescent periods with little or no
variability nor broad lines (Aretxaga \& Terlevich 1994).

\subsection{compact Supernova Remants}

\label{sec:cSNRs}

cSNRs are the key ingreedient in the starburst model. They are in fact
the only thing which set AGN-like starbursts apart from ordinary ones.
In the context of the starburst model for AGN, cSNRs are responsible
for the time variability, X-ray and radio emission, the BLR properties
and the NLR ionization, as explained in the seminal paper by Terlevich
\et (1992). cSNRs are nothing but ordinary SNRs evolving in a dense
circumstellar medium ($n_{CSM} > 10^6$~\percc). Density is the main
parameter governing the evolution of the remnant, though metallicity
and the detailed structure of ejecta and CSM also play a role. At such
high densities cooling of the shocked material is very efficient and
the remnant is capable of radiating away most of its kinetic energy in
a few years. The structure and evolution of a cSNR is complex and has
to be followed numerically, and hydro codes  have only recently been
adapted to work under such extreme conditions. The first simulations of
cSNRs were those by Terlevich \et (1992). More recent and detailed
calculations can be found in Plewa (1995), Terlevich \et (1995) and Cid
Fernandes \et (1996). The state of the art in numerical modeling of
radiative shocks is reviewed in T. Plewa's contribution.

To first order, the structure of a cSNR comprises 4 regions (see
schemes in Terlevich \et 1992 and Cid Fernandes \& Terlevich 1994).
From the inside out: (1) the unshocked, freely expanding ejecta, (2)
shocked ejecta, (3) shocked CSM and (4) unperturbed CSM. Dense,
cold, fast moving thin shells form behind both the forward and
reverse shocks via catastrophic cooling. These two cold regions plus
the unshocked ejecta are photoionized by the high energy photons
emanating from the cooling regions, producing broad emission lines
whose ratios, luminosities and widths are similar to those found in
AGN. Though many details remain to be worked
out, the cSNR model is definitely in the right track, as confirmed
by the discovery of objects like SN 1987F (Filippenko 1989), with a
spectrum so similar to that of AGN that led Filippenko to call it
a `Seyfert 1 impostor'. 

Besides being fascinating objects by themselves, cSNRs are 
potentially the true missing link between starbursts and AGN.
Whether or not they can explain AGN properties which have puzzled us
for so long, there can be little doubt that a starburst whose SNe
evolve into cSNRs will look like an AGN in many aspects. In my view,
this has been already {\it proven} both theoretically, with the work
described above, and empirically, with the discovery of `Seyfert 1
impostors'. At this stage, I cannot help observing that, unlike the
situation in AGN studies, the theory of cSNRs is parsecs ahead of
observations. cSNR simulations have reached an unprecedent stage of
refinement. We can, for instance, consider the detailed structure of
both ejecta and CSM, as well as the effects of inhomogeneities on
the evolution of the remnant. The theory here is really crying out to
be tested. Detailed observations of cSNRs are badly needed not only to
better guide the theory of radiative shocks but to strongly test the
cSNR-AGN connection proposed by the starburst model.

\subsection{Why high density?}

\label{sec:density}

All that is required for a SN to become a cSNR is a dense CSM. Given
this, a starburst will {\it certainly} present many AGN properties.
But what generates such a high density? The historical explanation
was based on the effects of metalicity (Terlevich \& Melnick 1985,
Terlevich, Melnick \& Moles 1987). The central regions of a galaxy,
particularly early type ones, where most AGN are found, have an
enhanced metallicity, which would enhance radiatively driven mass
loss, creating both warmers and a dense CSM around the cSNR
progenitors. This reasoning was used by Terlevich, Melnick \& Moles
to explain the predominance of AGN at early Hubble types and
starbursts at late types.

Though metallicity certainly plays some role, the current view is that
the ISM in a nuclear starburst is pressurized by stellar winds and
SNRs. The high pressure confines the wind driven bubbles to small
volumes, creating the conditions for SNe to become cSNRs (see J.
Franco's contribution elsewhere in this volume). This, however, is
probably not the only way to produce cSNRs, since objects like SN 1987F
and SN 1988Z (Filippenko 1989, Stathakis \& Sadler 1991) were found in
regions showing no signs of particularly violent star-formation.

\section{AGN properties explained by the model}

Having gone through the basics, I now briefly outline some of the
AGN phenomenology which can be at least reasonably well understood in
the the framework of the starburst model.

\subsection{BLR properties}

That cSNRs can have emision line regions aking to those of Seyfert 1
was already realized by Fransson (1988) in his early study of SN-CSM
interaction. Indeed, the combined hydrodynamics $+$\ photoionization
calculations of Terlevich \et (1992) demonstrated that cSNRs do at
least as well as canonical models in reproducing the BLR line ratios.
Furthermore the different photoionized regions (the two thin shell
associated with the forward and reverse shocks plus the ejecta) explain
the existence of high and low ionization line regions (e.g.,
Collin-Souffrin \et 1988). Reproducing the basic BLR properties with a
{\it physical} model, as opposed to an {\it ad hoc} model such as a
central ionizing source plus an arbitrary distribution of clouds, is
one of the main achievements of the starburst model.

Though the velocities and luminosities of the line emitting regions in
cSNRs are similar to those in the BLR, detailed emission line profile
calculations  proved to be more complicated than initially thought (Cid
Fernandes \& Terlevich 1994, Cid Fernandes 1995). The complications
arise because the line emitting shells are thought to be optically
thick, but geometrically too thin to be modeled with the Sobolev
approximation. In this situation calculations are very sensitive to the
detailed shape of the shells, which is likely to be distorted by
instabilities. Though the theory here still needs work, one can always
resort to the empirical argument that {\it observed} profiles in cSNRs
such as SN 1987F and SN 1988Z would probably go un-noticed if put among
a gallery of AGN profiles. A further complication is that, discounting
low luminosity systems ($M_B$ $\gapprox$-20), chances are that we
seldom observe isolated cSNRs in AGN. Interestingly, computations of
the line profiles for multi-cSNR systems yielded more robust results
than for individual cSNRs. Effects such as the line width-luminosity
correlation and the centrally depressed \Hb/\Ha\ profile ratio can be
readily understood in the model.

\subsection{Optical--UV variability and the origin of the lag}

Most of the recent interest in AGN variability has been on the time-lag
between continuum and emission line variations, with intensive
monitoring campaings designed to reconstruct the BLR geometry using
echo-mapping techniques. In the starburst model for AGN the lag is {\it
not} due to reverberation, but due to the hydrodynamical effects
associated with thin shell formation behind radiative shocks (Terlevich
\et 1995, Plewa 1995). As the shocked material begins to cool in the
processes leading to shell formation, lots of radiation are released,
producing a continuum burst. Eventually the cold gas collapses onto a
thin shell, but this only happens some time after the maximum in the
continuum, explaining the origin of the lag. As the continuum burst
fades and the shell density increases, the ionization parameter goes
down, explaining why low ionization lines take longer to respond to
continuum variations. After shell formation, shock oscillations produce
further variations, but this time the shell is already formed and there
should be little or no lag, explaining the puzzling observation that
some events show essentialy no lag (e.g., Clavel \et 1991). This
elegant, albeit controversial, interpretation illustrates the richness
of phenomena associated with cSNRs and radiative shocks in general.

The optical-UV continuum variability of starburst powered AGN is
reviewed in I. Aretxaga's contribution, where the reader can find how
properties such as the light curve statistics of Seyfert 1s, the
anti-correlation between variability and luminosity and the Structure
Function of QSOs are explained in the starburst model.

\subsection{QSO luminosity function}

Terlevich \& Boyle (1993) carried out the interesting exercise of
evolving the stars in the core of present day elliptical galaxies
back in time (see also Terlevich 1992 and Boyle 1994). With simple
assumptions they were able to reproduce the shape, amplitude and
evolution of the QSO luminosity function, supporting
not only the starburst model version of QSOs as young/primeval
galaxies but also galaxy formation models which indicate that giant
ellipticals have gone through a period of intense star-forming
activity around $z$~$\gapprox$~2. (See Heckman 1994 and Terlevich
1994 for an interesting debate on this issue.)

\subsection{Etcetera}

The points above are those which have been studied in more detail, but
there are other AGN properties which can also be understood in the
framework of the starburst model. These include (in varying stages of
development): the Ca~II absorption triplet (Terlevich, D\'{\i}az \&
Terlevich 1990), the spectral energy distribution (Terlevich 1990),
radio emission and variability (Colina 1993), the nature of strong
Fe~II emitters, size of the cluster and mass segregation (Terlevich
1994), effects of magnectic fields (R\'o\.zyczka \& Tenorio-Tagle 1995)
and X-ray features such as the warm-absorber, cold reflector and
high-energy cut-off.

\section{AGN properties not explained by the model}

Despite this impressive list of explainable properties, the
starburst model has from its early days been subjected to criticism
from several fronts. The similarities between radio quiet and radio
loud AGN, for instance, have always been a matter of worry, since
radio loud objects have from the outset been excluded by the model.
Other unresolved issues include micro-variability, and the absence
of a Lyman edge and stellar features in the UV. However, rapid X-ray
variability and the recently discovered broad Fe lines are, in my
view, the most serious dificulties currently faced by the model.

Rapid X-ray variability has always been seen as a serious problem for
non-black hole models for AGN, a point not so much based on theory, but
on the fact that galactic black-hole candidates (Cygnus X-1 being the
classic example) also show rapid variability (see Mushotsky, Done \&
Pounds 1993). Similarity arguments, however, have to be taken with
care, since on optical wavelengths (at least) cSNRs are much more
AGN-like than galactic black holes. Furthermore, gamma ray observations
are revealing a clear difference between the high energy spectra of
Cygnus X-1, which goes well into the MeV range, and that of radio-quiet
AGN, which turns-over at about 100 KeV and cuts off at a few hundred
KeV (A. Carrami\~nana, priv.\ comm., McConnell \et 1994, Warwick \et
1996). (See also Kinney 1994 for a comparison between stellar
accretion disk systems and AGN.) In any case, it is difficult to see
how cSNRs could produce strong, rapid X-ray variability. The
interaction of SN fragments with the dense, thin shells in cSNRs can
give rise to rapid X-ray flares, but, as discussed by Cid Fernandes \et
(1996), the physical conditions required to model AGN-like flares
quantitatively are too extreme. This, however, has yet to be verified
with X-ray monitoring of cSNRs---the fact that we don't know how to
produce strong, rapid variability does not mean that they don't have
it!

The discovery of an extremely broad ($\sim c/3$) Fe line in MGC
6-30-15 (Tanaka \et al 1995), a feature now detected in many other
Seyfert 1s (see R. Mushotsky contribution), brought further problems
for the starburst model. Besides being very difficult to see how
such a wide line could be produced in cSNRs, the observed profile is
remarkably well fit by a relativistic accretion disk model. Further
compelling evidence for a disk surrounding a compact super-massive
object came with the water maser observations of NGC 4258 by Miyoshi
\et (1995).

\section{Discussion: Starbursts, black-holes or both?}

All in all, the starburst model proved capable of explaining
several, though not all, AGN properties. What makes the model so
attractive is its deductive character, deriving observables out of
physics instead of parameters. Though the numerical balance of
evidence for and against may favour the starburst model, a single
unexplained observation is enough to put any model in trouble. When
an astrophysical model is in trouble we either (1) drop it
altogether, (2) modify/fix it or (3) opt to say it does not apply to
all objects. As already said above,  starbursts containing cSNRs
would certainly look like AGN in many ways, so there must be some,
maybe many objects of this kind in AGN lists. In this sense, at
least, ruling out a starburst origin for the activity in galactic
nuclei is not a sensible alternative. The choice therefore has to be
between fixing the model or restricting its applicability to a
sub-set of the AGN family.

The easiest way to account for rapid X-ray variability, broad iron line
and radio loud objects would be to go for a hybrid black-hole
$+$\ starburst scenario. One version of such a model is that of J.
Perry and colaborators, reviewed elsewhere in this volume. Naively, one
could think that combining a ``Terlevich-type'' of starburst with a
black-hole would fix the problems with X-rays, with stars and cSNRs
still responsible for the activity at lower energies. While this might
work in some cases, my view is that such a scheme does not provide a
natural explanation for the correlations between high and low energy
properties of AGN, one example being the correlation between X-ray and
Balmer line luminosities (e.g., Ward \et 1988). Such correlations
indicate that the low and high energy photons somehow know about each
other, something which would be difficult to understand if they
originate from very different phenomena. It is therefore not obvious
that a simple combination of the canonical and starburst paradigms is
the answer.

I conclude that, to the disgust of purists, the evidence seems to be
pointing to two kinds of objects: starburst and accretion-powered
AGN. This is not a new idea, as many of us always suspected that AGN
constitute a mixed bag. Distinguisuing
between these two possibilities has been a central theme in the
starburst $\times$\ monster debate (see Filippenko \et 1993). At
this stage, detection of strong, rapid X-ray variability and broad
Fe lines are the best indication of the existence an accretion-disk
in AGN. Experiments to resolve the small but extended nucleus
predicted by the starburst model will eventually provide a more
conclusive method to discriminate starburst from accretion-powered
AGN.

Going back to our original question, can we after all these years
reach a veredictum on whether starbursts can power AGN? Given the
advances on cSNR theory and the discovery of objects like SN 1987F
and SN 1988Z, we can safely say that yes, starbursts {\it can} power
AGN. At the same time, recent discoveries have put the standard
accretion-disk model on firm ground. Whatever the final answer to
this cosmic puzzle is, it is clear that while we digest this
apparent conflict, much can be learned observing cSNRs. A
fundamental aspect which differentiates the starburst model from 
black-hole and hybrid models is that many of its predictions can be
tested {\it outside} AGN. Galactic black-hole candidates are about
the closest one can get to an AGN-like engine, but applying the
knowledge of such systems to AGN involves an unconfortable leap of
several factors of 10. cSNRs, on the contrary, are expected and
observed to be similar to those inferred to exist in the cores of
massive starburst-powered AGN. They therefore provide a unique
laboratory to strongly test the starburst-AGN connection and to help
us solve this yet unfinished debate.

%RRRRRRRRRRRRRRRRRRRRRRRRRRRRRRRRRRRRRRRRRRRRRRRRRRRRRRRRRRRRRRRRRRRRRRR
%				REFERENCES
%RRRRRRRRRRRRRRRRRRRRRRRRRRRRRRRRRRRRRRRRRRRRRRRRRRRRRRRRRRRRRRRRRRRRRRR

%RRRRRRRRRRRRRRRRRRRRRRRRRRRRRRRRRRRRRRRRRRRRRRRRRRRRRRRRRRRRRRRRRRRRRRR

\end{document}